# SPRING: Speech and PRonunciation ImprovemeNt through Games, for Hispanic children

Anuj Tewari, Nitesh Goyal, Matthew K. Chan, Tina Yau, John Canny and Ulrik Schroeder

*Abstract*— **Lack of proper English pronunciations is a major problem for immigrant population in developed countries like U.S. This poses various problems, including a barrier to entry into mainstream society. This paper presents a research study that explores the use of speech technologies merged with activity-based and arcade-based games to do pronunciation feedback for Hispanic children within the U.S. A 3-month long study with immigrant population in California was used to investigate and analyze the effectiveness of computer aided pronunciation feedback through games. In addition to quantitative findings that point to statistically significant gains in pronunciation quality, the paper also explores qualitative findings, interaction patterns and challenges faced by the researchers in dealing with this community. It also describes the issues involved in dealing with pronunciation as a competency.**

*Index Terms*—**Developed regions, Education, Hispanic, Educational technology, Games, Pronunciation**

## I. INTRODUCTION

THIS paper presents a rare effort and a rare approach towards English language learning. It is a rare effort because the focus of the ICTD community has long been the developing parts of the world as opposed to the underprivileged or under-resourced children in the developed world. This paper has a rare approach because it investigates the effectiveness and viability of use of speech technologies and games resembling hugely popular commercial games (which have not been used together in the past) in giving pronunciation feedback for improving English pronunciation of Hispanic immigrant children.

"Inclusive Education" is a part of *Improving Education Quality,* one of the themes of UNESCO. Children belonging to indigenous groups and linguistic minorities are classified as vulnerable to exclusion from the benefits of the education system. Traditionally such minorities have been believed to exist only in the developing and the less developed world. However, statistics and our experiences suggest that such minorities exist even in the developed world. International Bureau of Education (IBE), an international centre for the content of education, is an integral, yet autonomous part of UNESCO and International Academy of Education (IAE). IBE's *Teaching Additional Languages* booklet classifies "speaking" as an integral part of language learning for additional language learners.

As a contribution to include such minorities and address the challenges involved, we describe a three month long study with Hispanic immigrant children with limited exposure to spoken English language at a public high school to explore the potential role of pronunciation-feedback coupled games as motivational tools, henceforth referred to as SPRING, to teach and improve pronunciation of immigrant children.

Around 6 percent of the total population in USA is of Mexican origin – authorized or not [1][2]. About 70 percent of these Mexican born immigrants live in closed communities in just four of the fifty states in USA: California, Texas, Illinois, and Arizona [3]. These communities do not just live together for cultural and social benefits. Their similar economic and financial conditions also bring them closer because the Spanish-speaking immigrant population is almost twice as likely to live in poverty, much higher compared to any other immigrant group [3]. According to MPI's release in February 2010, about three-quarters of Mexican immigrants in 2008 were limited English proficient.

This highlights the plight of Hispanic, and specifically Mexican, immigrant cultural and linguistic minority living in USA, one of the most developed countries. Evidently, this community suffers from exclusion of benefits of the infrastructure and society available in the developed world. Moreover, their lack of knowledge of primary language of communication: English hampers prospects of improvement.

Thus, our study focuses on the age group of 12-18 year old immigrant Hispanic children by employing games similar to the games that they already enjoy playing as an aid to the existing classroom teaching in English Language Learners (ELL) classes.

The next section, Literature Review (II) refers to similar conducted research by the community and explains what sets our work apart from the existing works. Section Study (IV) describers the study locale and setup. Section IX describes the Quantitative Results obtained from the post study interviews.

Manuscript received April 2, 2010. This was supported in parts by Verizon Wireless.

Anuj Tewari is a graduate student in the department of Electrical Engineering and Computer Science (EECS) (phone: +1 312-613-4959; email: anuj@eecs.berkeley.edu). John Canny is the founding director of BID and holds the Paul and Stacy Jacobs Distinguished Professorship of Engineering at UC Berkeley.

Nitesh Goyal is a visiting research scholar at the Berkeley Institute of Design (BID). Matthew Chan and Tina Yau are undergrads in EECS, who are also affiliated with BID.

This is followed by the Qualitative Results in Section X. We discuss some of the important and interesting challenges in Section XI followed by the Conclusion and Future Works in Section XII.

## II. RELATED WORK

Computer Assisted Language Learning (CALL) has existed for almost 70 years now. Several methods and systems have been proposed to help improve particular focus areas in language learning using computers. Most work in the CALL domain does not explore the ability of technology to teach English pronunciation using persuasive computer games to immigrant high school children.

Horowitz et al. [4] describes an 8-week long study that promotes literacy in USA with participants from households below the poverty line. The focus of this study was to improve literacy and teach the English alphabet using videos. While the videos were persuasive, they lacked focus on improving the English pronunciation and were targeted at very young children.

Massaro in 2003 [5] described Baldy, a virtual talking head on a screen with focus on helping users learn how to pronounce the phonemes properly as a virtual teacher. This is by far the only study we know that focuses on teaching pronunciation and provided a visually detailed feedback and training. Powers et al [6] is also a similar system and goes a step further by acting as Embodied Conversational Agents (ECA). These systems improve upon Massaro by including other features like vocabulary learning etc. While these systems mention encouraging results, they lack information about how motivational these systems might be.

Multimodality has also been briefly investigated for pedagogical benefits in English Language Learning. Chen Yu et al [7] suggests that spoken language can be grounded sensory perceptions of the real world. It describes a learning interface that bridges a gap between the real world physical objects and the virtual interface. Sluis et al [8] describes a collaborative table top based simple matching to help develop the reading skills of young groups of children. Fallahkhair et al [9] describes a system with 2 inter coupled-interfaces: TV as an audio visual aid, and mobile phone as a supporting aid to help learners learn the vocabulary. These systems also continue to focus on the writing, reading, and vocabulary parts of the language.

However, recently there has been a growing interest in including computer based tools that use automated speech recognition to provide a guided reading experience for the users. Mostow et al's Project LISTEN based Reading Tutor [10][11][12] has been used with a variety of audiences in improving the English reading ability of children, with English as a first language and with English as a second language (ESL/ELL) in USA by Poulsen et al [13] and Canada by Reeder et al [14].

While the Reading Tutor involves use of stories, Kam et al [15] has successfully shown the use of games, especially the use of mobile games as persuasive tools for improving the English literacy of the illiterate English as Second Language (ESL/ELL) children in India. Johnson et al in 2005[16] and 2007 [17] present a system being used by the US Army to learn Arabic in Iraq. However, we feel that due to nature of the intended use and lack of a particular pronunciation focus, this product is unsuitable for use by young children.

Anna et al's DEAL[18] uses both the ECA and task-based game design in its system. The users hence, learn how to structure the sentences properly and learn appropriate word placements. This system is focused more on the grammar than spoken language.

As explained above, the existing works successfully describe using games or speech recognition or both for literacy improvement. However, unlike our system, none of them employs usage of both games and speech recognition for pronunciation improvement amongst English as Language Learner (ELL) children.

## III. OVERVIEW OF PILOT STUDY

The pilot study was carried out at a public high school located in a highly populated Hispanic immigrant location in California, USA for about three months from December, 2009 to March, 2010. The study took place within the school premises during the extended school timings and involved demographic study, the pre-test, the experiment, and the post-test with permissions from the school authorities, the teachers involved, the students and/or their parents.

Three sessions were held, on an average, per week for four weeks. Each session accommodated approximately three students, one after the other. So, each student played freely in seclusion from the other students for about ten minutes per week. There were two different games that each student was able to play. These games were alternated each week to keep up the interest level of the students. Hence, during the four week long endeavor, each student received a total of about forty minutes worth of play time from the two games of SPRING.

## IV. STUDY LOCALE AND SETUP

This section describes the steps we followed to find our user group. We began by contacting the teachers and school authorities at several public middle and high schools located in the vicinity. Our aim was to locate a school with a high immigrant population having a low level of spoken English fluency. Based upon the anonymous demographic and diversity data that we received from these schools, we shortlisted three schools where we submitted a request for conducting research with the students within the school premises. One of the public High Schools that accepted our request fulfilled our requirements.

According to a data survey by the school district in 2007-08, an overall 50% attrition for ELL was reported. For this particular High School, the rate was 75% for ELL. We were guided to one of the English Language Learner's (ELL) classes at this school. The class consisted of 20 students, at ELL level 2. These students had been in USA for less than two years, and had over the time attended, and cleared ELL level

1. The class had a 100 percent immigrant population. In this class, 90% immigrants were from Latin America. Of the students in this class, 95% are labeled as SED (Socio-Economically Disadvantaged). That means one of two things (or both things) is true of all but one student. Either (1) the students are living at an economic level qualifying them for the federal free/reduced lunch program or (2) his/her parents did not graduate from high school, or both are true.

This situation at this school compares favorably with the previously quoted national data. So, we decided to choose this particular ELL class at this school.

## V. Data Collection

The pilot study was managed solely by the four researchers involved in this project. However, due to the nature of the participants, a local member of the school volunteered to help translate between English and Spanish for children who could not understand our use of English language.

The class had a strength of 20 students. We divided this class, for the purpose of our study, into two groups of 10 students each. One of the groups was the CONTROL GROUP, which received the regular classroom training from the teacher and did not attend the play sessions with SPRING. The other half, EXPERIMENT GROUP, received exactly the same training in the classroom as the CONTROL GROUP. However, they also received the opportunity to attend play sessions with SPRING.

To reduce any bias due to pre-existing knowledge between the two groups, we randomly picked and assigned the students to either of the two groups. Next, we administered a simple qualifying test to all the 20 participants to gather their existing level of knowledge. The test consisted of a slideshow of 30 words, one after the other, on a computer. The test taker was required to speak the word shown on the screen and a speech recognition engine (discussed later) recorded and scored the utterance. The scores were not made visible to the students to reduce anxiety. The tests were done in private with each student to minimize any learning effects. The words selected for the test were kept constant for the entire pool of the participants and were selected from the syllabus and the recommended textbook for that class. These sessions were also audio recorded. During the course of the study, we evaluated the participants using a similar test to prevent test anxiety and for consistent comparable results. These were administered as a series of pre-tests and post-tests.

Each play session with SPRING was video taped to record the emotional state of the participants while playing. This was captured by facial and body expressions, exclamations, sighs, gasps and other auditory feedback. These recordings created the contextual data by providing us with more data about the playability of the different stages, elements and parts of our games.

## VI. Participants

This pilot study was one of the first kinds to be established at our partner school, especially with the immigrant population. So, our participants were very new to this new arrangement and we benefitted from their enthusiasm to participate in "something new". Initially, in total we obtained consent from 20 children and/or their parents to participate in the study. They were all part of the same ELL Grade 2 class at the school and represented the total strength of the class, as well. We began our pilot study with all the 20 of them. However, during the due course of time, 2 of them left the study. Unfortunately, the reasons for attrition could not be conclusively determined due to their continuous absence from the school itself during the three month long duration. However, reasons of attrition, after consultation with teacher, seemed to be family and financial problems for the male participant, and teen-age pregnancy for the girl participant.

### A. Demographics

The 18 students (after attrition of 2 from 20) exhibited the following characteristics:
- Six (6) were male and twelve (12) were females.
- All eighteen (18) in the study were in ELL level 2.
- The students were in the age range of 14 to 17.
- All eighteen (18) students were of Hispanic ethnicity

Many of them lived with family members such as uncles, aunts, and cousins; some did not live with their mothers or fathers.

The fathers, uncles, and brothers held jobs working in a market, as a florist, washing cars, as a gardener, or other lower-end jobs. Few had younger/older brothers or sisters still in school.

The mothers, aunts, and sisters had jobs that involved cleaning homes, babysitting, or no job at all.

Amongst the 18 children, many had ambitions of becoming a lawyer/attorney, doctors, teacher etc. 16 of the 18 students either had a cell phone or had access to a cell phone (from a family member) and only use it for texting or talking on the phone; none play the games on the phone. When asked about what kind of games they played, students listed board games such as checkers to several Playstation games such as soccer (FIFA), Boxing, racing games, Mario, or some computer games. There were a small number of students who didn't play games at all, too. None of them knew about Guitar Hero.

When it comes to learning English, all the students pointed out vocabulary acquisition and pronunciation/speaking as their key issues; other issues were reading and writing. All the students except one recognize the importance of learning of English, so they can attain better job prospects and communicate better. However, the teachers also mentioned that there is some resistance to learning English because these students are surrounded by a community of other Spanish-speaking peers and lowers their incentive to learn. These students also mentioned peer pressure because they did not want to sound silly when they mispronounce English words. Evidently, there are issues with intrinsic motivation.

While lack of intrinsic motivation is a discouraging factor, the extrinsic motivation is also lacking. While the children want to succeed and aim high for their life, there are not many good examples available in their community. Furthermore, for illegal immigrants, avenues for higher education and professional growth are virtually non-existent. This reduces the motivation of some of the students to try harder because they know that they will eventually get low skilled and low-waged jobs like their parents.

VII. DESIGN

This sections describes how we designed our study and the associated apparatus and content for a successful implementation. We begin by explaining the current curriculum taught at the school to our user group and how we derived a syllabus for our study. Next we explain the methodology behind our game designs and end with a description of the implementation and system design.

*A. Curriculum Design*

A student in the ELL Level 2 spends roughly 3 hours in the ELL classroom daily. This includes instruction and teaching, drills, practice sessions, silent readings, and tutor-time. We developed our curriculum worth teaching 7 percent of the entire vocabulary, for the entire academic year, in about 10 minutes session once a week after discussing with the ELL teacher for the class. This represents quite a negligible self-learning time. The students at ELL Level 2 at the chosen school attend classroom teaching by an experienced teacher, aided by audio-visual media to improve the attention and understanding. They follow the curriculum designed according to the textbook "Milestones California Edition". The book is divided into six units, each describing a different facet of life like "Dreams", and "Survival" etc. This curriculum is heavily based on reading, vocabulary, and grammar lessons in content, and exercises. However, it offers limited opportunities to speak English formally. Each ELL level requires a certain minimum level of knowledge of English vocabulary. These words are discussed in the class but spoken & pronunciation correction drills of these words do not happen at the class or an individual level. The only opportunity that these children have at listening these words are when used by the teacher in the class during the discussions.

The "Milestones" book includes a list of about 300 words from the 6 units that are expected to be known by the students at the end of the academic year. We divided these 6 units into 3 parts: Group A: Units 1 and 2 which had been taught by the teacher in the class before we began the study; Group B: Units 3 and 4 which were being taught during the study; Group C: Units 5 and 6 which had not been taught during the duration of the study. We randomly chose 10 words out of each Group (A, B, and C), giving 30 words, a 10% sample set out of the pool of 300 words and created a syllabus of our study based on them. The aim to divide the words into the groups was to investigate if the games caused significant deviation between the learning gains of pre-existing knowledge (Group A), or unknown knowledge (Group C), or aided what is being taught (Group B). Some of the words in the sample set included "Menacing", "Attic" and "Soggy" etc.

The study was designed to test the pronunciation ability of this sample set of words by the users, teach the users how to pronounce those words using a game, and then finally testing to detect the effects, if any.

*B. System Design*

The entire game logic for SPRING was written in Flash Actionscript. SPRING was eventually deployed on an Ubuntu Linux installation. Details of the individual pieces are as follow:

*1) Speech recognizer*

For the purposes of the speech recognition, we used the CMU Sphinx-III speech recognition engine. However, instead of using it in decoding mode we used it in forced-alignment mode. In force-alignment, rather than being given a set of possible words to search for, the search engine in the recognizer is given an exact transcription of what is being spoken in the speech data. A reason for using the force-alignment mode was that we were able to obtain scores at the level of individual phonetic units. We used this information to point out which part of a particular word was uttered incorrectly.

*2) Speaker adaptation*

Since we wanted the games to give feedback after comparing to standard American accented pronunciations, we trained the recognizer on large corpuses of data (15GB, raw format) from American accented speakers. However, we did account for the change in texture from a male to female voice. We recorded audio utterances from 2 American males and 2 American females and used MLLR (Maximum Likelihood Linear Regression) transforms to adapt the recognizer to male and female voices as and when required. Use of MLLR transforms is the most commonly used method for speaker adaptation in automatic speech recognition systems.

*3) Feedback routine*

The recognizer could generate acoustic scores, but they had to be compared against standard American accented pronunciations, before giving feedback to the participants on how they did on a particular word utterance. Therefore, we coded a library that returned back a Likert scale (1-3) rating for each phonetic unit in the word under consideration. This rating could then be used to give feedback to the participant.

*4) Graduated interval recall*

The game logic for Voz.Guitar was implemented in a way that the syllabus queue was chosen according to a well-established algorithm called Graduated Interval Recall. [19] The algorithm helps in determining the order of the questions, given a syllabus. It is modeled in a way that performance on a particular question determines the number of times it will be posed in the near future, thereby causing long-term retention of syllabus items. The game concept of Voz.Guitar had an aspect of repetition, as opposed to Zorro, which allowed the player to explore an exciting but static and pre-defined game world. Therefore we just used the algorithm for Voz.Guitar and not for Zorro. However, we countered the lack of repetition in Zorro, by making the participants play the game again. Moreover, we ensured that the time for which the participants are getting instructed (also playing the game) stays constant across the two games.

*C. Game Design*

The aim of the study was to design and create games, enriched with pedagogy that might motivate the players to play them, despite the challenges posed by the learning material in the game. We based the design of our games on the following resources:

1) Demographic interviews of the children clearly indicated a penchant for certain types of games.
2) Popular and best selling commercial software available in the market.

This gave us an advantage of creating games that were likely to peak interest of the children while they incorporated the best practices of game design and elements from existing games. Using this knowledge, we decided to create two games: Zorro (based on Mario), and Voz.Guitar (based on Guitar Hero) for chiefly the following reasons:

1) Activity based vs. Arcade based: The demographic interviews pointed out the predilection for two different genres: card games and action games. However, in either genre, the children preferred fast paced, non-time restricted gaming sessions.
2) Novel vs. Comfortable: We based our design on two popular and proven games — Mario and Guitar Hero. The demographic interviews indicated the previous playing experience of most of the participants with Mario, while none knew about Guitar Hero. So, we decided to give them a mix of a comforting known game and a novel, and hopefully exciting, game.
3) Adaptive vs. Non Adaptive: We chose Mario based game because it is non-adaptive and gives a consistent experience of play, with onus on the player to act fast. On the other hand, Guitar Hero based game was adaptive and had an element of surprise.

Both games followed the principle of teaching, drill, immediate feedback, scores, and repetition. Both games feature the word, associated playable American accented female voice, and spelled-out-pronunciation to aid the users. The spelled-out-pronunciations were obtained from the online dictionaries [20] and then modified accordingly by a trained linguist with five years of experience.

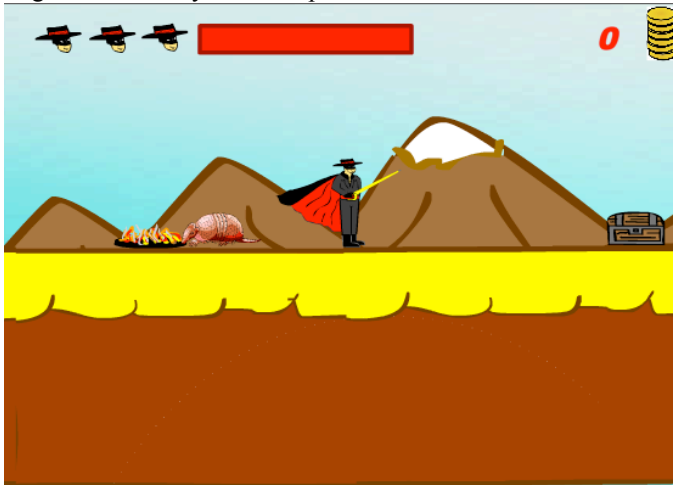

Figure 1

Zorro, as shown in figure [1], is a character based arcade game, which involves moving Zorro, the main character, of the game from left to the right of the scene using arrow keys until he reaches the castle.

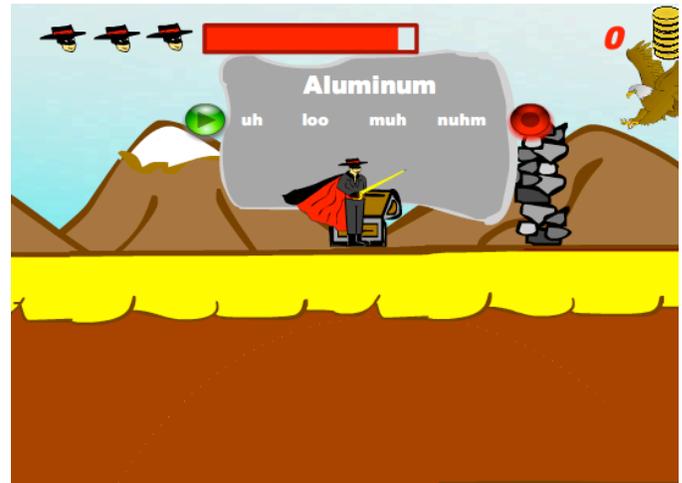

Figure 2

As shown in figure [2], on the way, he encounters five closed chests, dangerous animals, tricky terrain, and obstacles, which must be overcome. The obstacles can only be overcome by opening up the chests. Each chest contains a word, associated pronunciation, and the associated audio pronunciation coupled animated spelled-out-pronunciation. The word is pronounced three times every time it is played. Next, the user gets an opportunity to record their pronunciation of the word by the click of a button.

As shown in Figure [3], a feedback screen shows the correct and wrong parts of the pronunciation, and the associated score follows this. She also hears her own pronunciation and the intended pronunciation. After crossing the five obstacles by practicing the five words and avoiding the deadly animals, the user wins the game. In case, she finished short of 10 minutes, she is obliged to play the game again.

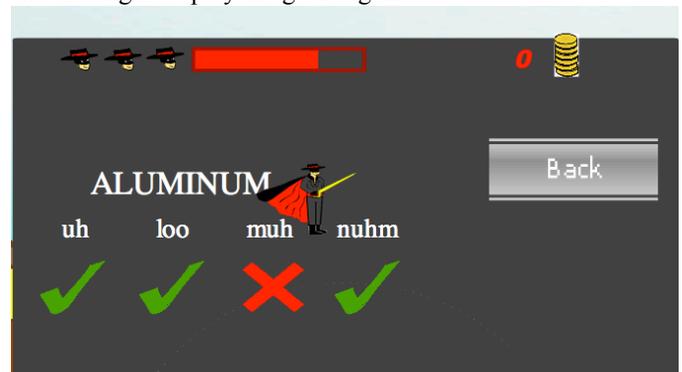

Figure 3

Voz.Guitar, as shown in Figure [4], is an activity-based game that displays the word, associated spelled-out pronunciation, and plays the associated pronunciation.

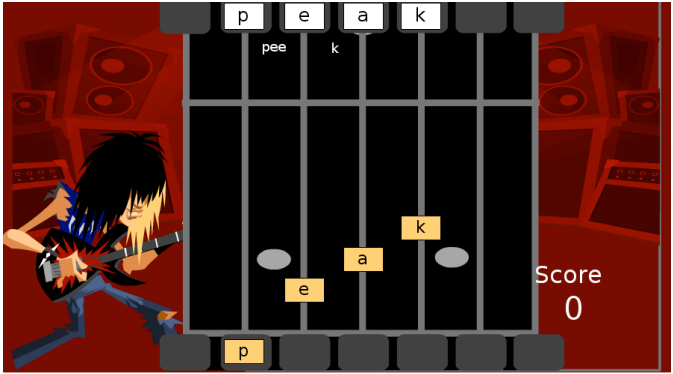

Figure 4

Next, it allows the users to hit the falling alphabets of the words at the right time. Next, the user is obliged to pronounce the word, as shown in Figure [5].

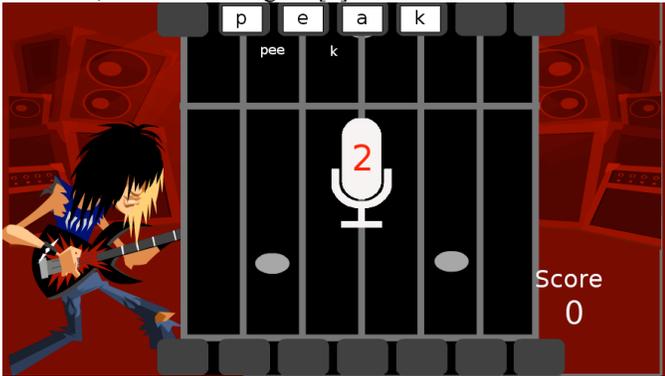

Figure 5

The feedback screen displays the hits and misses in the spelled-out-pronunciation and corresponding errors. The user hears her pronunciation followed by the intended pronunciation. The game is adaptive and hence, tends to automatically repeat the words, which have not received a satisfactory pronunciation response from the players, as shown in Figure [6].

Each positive utterance increases the score of the users. The session continues until the time limit of the session reaches.

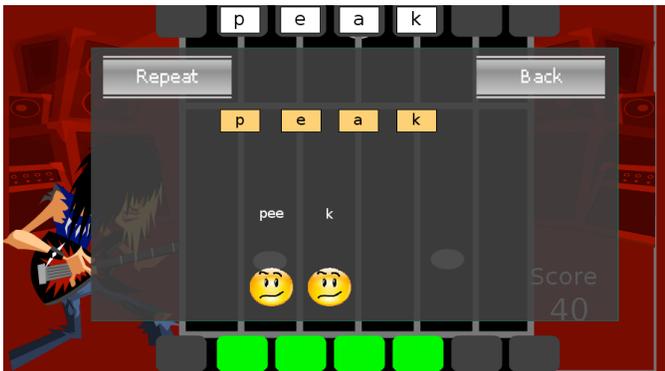

Figure 6

## VIII. STUDY SESSIONS

As previously mentioned, the study was designed across three groups of words, for two sample sets of population: Control Group, and Experimental Group.

The sessions lasted for around two hours per day, three times per week, and four weeks in a row. There were two types of sessions: Pre/Post Test sessions, and Learning sessions. A 2-hour Learning session was typically structured as follows: preparing the game database with the pre-determined group of words (A, or C), arrival at the school premises, arrangement and setup at a quiet location in one of the pre-arranged labs, greeting with the teacher, a list exchange of the students needed for that day, escorting a student to the lab, explanation of how the game is played, the goals, and a demonstration, the gaming/learning session for 10 minutes, a post game session qualitative interview, escorted return of the student, and bringing back the next student. A Pre/Post Test session involved the same as above except the student faced the test instead of the gaming session.

TABLE I
ASGP FOR CONTROL AND TREATMENT GROUPS

| Control Group | | Treatment Group | |
|---|---|---|---|
| CG1 | 1.24% | TG1 | 0.26% |
| CG2 | -1.72% | TG2 | 0.67% |
| CG3 | 1.15% | TG3 | 1.02% |
| CG4 | 0.71% | TG4 | -0.13% |
| CG5 | 1.02% | TG5 | 3.14% |
| CG6 | -7.40% | TG6 | 1.32% |
| CG7 | -1.68% | TG7 | -0.12% |
| CG8 | 0.71% | TG8 | 1.43% |
| CG9 | -0.12% | TG9 | 5.14% |

## IX. QUANTITATIVE OBSERVATIONS AND FINDINGS

### A. Metrics

Before we go on to explain our quantitative findings from the experiment, we need to define the metrics that we used to gauge the change in pronunciation. We used the following two metrics:

- Acoustic score gain percentages (ASGP): These were numerical scores generated by the CMU Sphinx-III speech recognizer. We did a batch decoding of all the audio samples (pre-test and post-test) that we had from the participants and generated acoustic scores to quantify the quality of pronunciation. The acoustic scores take all aspects of spoken language into account (like intonation, fluency, clarity etc). Moreover, the acoustic scores were generated for each phonetic unit in a word, and hence judge the actual quality of each phonetic unit. These individual phonetic unit level scores can be added together to generate word level scores. The ASGP for each participant was calculated as follows:

$$\frac{(TotalPoTS - TotalPrTS) * 100}{TotalPrTS}$$

Where TotalPoTS = Total Post-test scores and TotalPrTS=Total Pre-test scores.

- Word gain (WG): The word gain was nothing but the difference in the number of words that the recognizer could decode during the pre-test and the post-test. In simple words, this metric is a high-level representation of the number of words a participant learned to pronounce (with acceptable pronunciation) during the course of the experiment.

It should be noted that we had initially divided the 20 words in our curriculum (that was taught), into two parts. As explained earlier the first part came from pool of words they had already encountered in class and the second part came from pool of words that were completely unfamiliar to them. When we analyzed our post-test and pre-test data, we realized that the correlation between the category (familiar or unfamiliar) of the word and average gain on the same over the duration of the experiment was negligible. Quantitatively speaking, the correlations between the average scores (both ASGP and WG) across all participants and the category (familiar/unfamiliar) was <=0.27 for all the 20 words. Moreover, this was true for both, control and the treatment group. Hence we decided to group our results together, and analyze the gains across all the 20 words.

### B. Post-test gains

In each experiment, we used a standard statistical *t-test* to compare the gains of the treatment and the control group. This test yields a *p*-value indicating how significant the difference is between the means of the two groups. A two-tailed t-test on the pre-test scores of the treatment and the control group yielded a p-value of 0.25, which shows that there was not a statistically significant difference between the means of the two groups before the start of the experiment.

#### 1) Acoustic Score Gain Percentages (ASGP)

After the post-test, the mean acoustic score gain percentage for the control group was -0.68 ($\sigma$=2.77, n=9) and that for the treatment group was 1.41 ($\sigma$=1.72, n=9). The ASGP are small numbers because they are percentages of total pre-test scores across 20 words (more than 110 phonetic units). However, a two-tailed t-test between the ASGP for the control and the treatment group yielded a statistically significant p-value of 0.08. Table 1 lists the ASGP for participants in the control and treatment group.

A negative percentage denotes that the participant's total acoustic score for the post-test was lower than her acoustic score for the pre-test, and therefore the increase was actually negative.

#### 2) Word Gains (WG)

After the post-test, the word gain scores had a mean of 0 ($\sigma$=0.71, n=9) for the control group and a mean of 1.11 ($\sigma$=1.54, n=9) for the treatment group. This gain was in addition to the improvement in the quality of the pronunciations that is represented by the ASGP. Tables 2 and 3 list out the words attempted in pre-test, post-test and the resulting WG for the control and the treatment groups.

TABLE II
WG FOR CONTROL GROUP

| Participant ID | Number of words attempted in pre-test | Number of words attempted in post-test | WG |
| --- | --- | --- | --- |
| CG1 | 13 | 14 | 1.00 |
| CG2 | 15 | 15 | 0.00 |
| CG3 | 16 | 15 | -1.00 |
| CG4 | 12 | 12 | 0.00 |
| CG5 | 16 | 16 | 0.00 |
| CG6 | 17 | 16 | -1.00 |
| CG7 | 19 | 19 | 0.00 |
| CG8 | 12 | 12 | 0.00 |
| CG9 | 17 | 18 | 1.00 |

There wasn't a significant difference in the number of words the control and the treatment group could pronounce to some extent at the start of the experiment. The t-test on the number of words attempted at the start of the experiment yielded a value of 0.42.

However, the t-test on the number of words attempted at the end of the experiment (by the control and the treatment group) yielded a value of 0.06, which shows a statistically significant difference. It also points to a possible confidence boost during the study in terms of pronouncing less familiar and more complex words. Moreover, a two-tailed t-test on the WG values for the control and the treatment group yielded a p-value of 0.07. This shows that there was a statistically significant difference between the WG of the control and the treatment group.

TABLE III
WG FOR TREATMENT GROUP

| Participant ID | Number of words attempted in pre-test | Number of words attempted in post-test | WG |
| --- | --- | --- | --- |
| TG1 | 20 | 18 | -2.00 |
| TG2 | 19 | 20 | 1.00 |
| TG3 | 16 | 17 | 1.00 |
| TG4 | 15 | 16 | 1.00 |
| TG5 | 12 | 15 | 3.00 |
| TG6 | 13 | 14 | 1.00 |
| TG7 | 17 | 20 | 3.00 |
| TG8 | 19 | 19 | 0.00 |
| TG9 | 15 | 17 | 2.00 |

### C. Gender related findings

Our control and treatment group had the same distribution in terms of gender. Therefore, we also did some analysis to quantitatively measure the influence of gender on game play and learning.

The correlation between gender and ASGP for the control group (0.65) suggests that boys performed worse than the girls overall, over the period of the experiment. However, the correlation between gender and ASGP for the treatment group (0.32) suggests that gender did not influence the improvement in pronunciation quality that was exhibited by the participants, after playing the games. This is in contrast to the findings from similar ESL (English as a Second Language) acquisition studies in the more underprivileged parts of the world. [14]

### D. Effects of pre-test on post-test gains

The correlation between pre-test scores and ASGP for the treatment group was 0.11 and the correlation between pre-test scores and WG was 0.25. This shows that the participants in the study showed similar learning gains across both metrics irrespective of their performance on the pre-test. Therefore, there was no notion of bimodality as suggested by similar ESL acquisition studies in developing parts of the world. [15] This might be happening due to various different factors like better ESL levels, prior exposure to technology, and access to education.

### E. Learning gains during game play

We also collected data logs of how the participants performed during a session. Across a total of 10 (one of them dropped out of the school before the post-test) participants and a total of 40 game sessions the treatment group exhibited an average ASGP of 12%. Calculating the differences in acoustic scores of the first and last instance of a particular word in a single game session and averaging it across all participants in the treatment group resulted in these percentages.

## X. QUALITATIVE OBSERVATIONS AND FINDINGS

In addition to the quantitative sources of data, we also had videos that served as an important part of the analysis. We recorded approximately 600 minutes (10 hours of video). After transcription and qualitative coding of the data we came up with the following major qualitative findings:

### A. Player profiles

Through the duration of our study, we observed several key distinctions in our pool of subjects. The first major separation appeared in gender difference. The females appeared to be indifferent to the game play and were more focused on the speech/voice features of the game. Females also needed more assistance with the games compared to males, whether it were additional verbal cues or helping them with the obstacles in the game. When a translator was used for one female, the two put together were more engaged with playing both Zorro and Voz.Guitar; the two laughed, gestured and were more focused on game play in addition to the speech/voice features.

The males were more focused on game play than the speech/voice features; for example, when they opened the chest with Zorro and the feedback screen appeared, the males were still playing with the Zorro character (trying to move it around). When the males did interact with the speech/voice features, they said the words with more confidence than the females and had less stuttering and hesitation.

For both male and females, they exhibited a certain learning curve when playing, and that was true for both games. Almost none of the players used the "repeat" feature in the Zorro game, as opposed to Voz.Guitar that forced them by having them go through each word again. In addition, although both genders found the games entertaining as a whole, they did occasionally display gestures of frustration including rolling their eyes and hand waving to brush off mistakes. We felt that these gestures were partly attributed to general game playing

TABLE IV: PLAYER PROFILES: GAME DESIGN SUGGESTIONS

| Name | Sex Male/Female | Likes to play games | Body language while playing | Game Involvement | Pedagogy Involvement | Game Design Suggestions |
|---|---|---|---|---|---|---|
| Pablo | Male | Yes | Active, Focused, Nimble | High: Focused on Scores | Low, Bypasses the learning | Game requires higher percentage of accuracy to bypass the pedagogy elements |
| Juna | Female | Yes | Indifferent | Little: Takes a call on phone during the play | Low | Other Genre Games like Shopping Spree, Pop culture, Dressing up |
| Estera | Female | Yes | Not too excited | High: If game is socially interactive | OK | Online Social Networked Games with discussions, and chatting |
| Sandra | Female | No | Confused | Frustrated: Loses Lives constantly | OK | Games with easier levels, and abundant practice for learning game controls |

and demonstrate the student's attention and involvement in the game, which is a positive factor.

We further broke down our subject pool and found four specific player profile classifications in the subject population. They are represented by the following four names: **Pablo, Juna, Estera, Sandra**. And we suggest some design decisions to be kept in mind for future designs to create an inclusive game in Table IV.

### B. Pronunciation Measures

We tested and identified pronunciation measures using a speech recognizer. Since all the processing was happening off the field and on a dedicated machine, we got accurate scores. However, to bring more credibility and to add more human aspect to our research, we would like to seek help from trained linguists. Moreover, we would want to get Likert scale readings from general American population, like a housewife or a salesman at a supermarket. Our overarching goal is to better the pronunciations to a level that is socially acceptable. Using the recognizer for evaluation is the first step, but using human inputs from various different sources would be beneficial.

### C. Other Findings

In the post-game play session interview, 7 out of 9 participants reported that they felt they were learning pronunciations during the game, the rest said they did not know if they learned. We also asked who they would want to help them with pronunciations. 6 out of 9 participants said they would want help from both their teacher and SPRING. The rest of the 3 participants said they would want to learn from the game only. Since this is self-reported data, we don't attach a lot of value to it, but it definitely points out that SPRING was a pleasant change for a majority of the students. When asked which game they enjoyed more, 6 out of 9 said they liked Zorro better than Voz.Guitar, the rest of the 3 said the opposite. This was intuitive because Voz.Guitar had a lot of repetition and Zorro was exciting. We would want to mix these two factors in the next phases of the study. It would have been hard to mix game play and pedagogical concepts right from the start, but now we can use the current phase of study and the design decisions we took to inform the next phase of the study and design.

## XI. CHALLENGES FACED

### A. Motivation

The community we worked with was a very complex one. There was little or no motivation for them to acquire English as a Second Language. Through our games, we were trying to break this barrier to entry. Our aim was to develop games that are inherently more engaging and have pedagogical concepts merged into game concepts. Throughout the duration of the study we constantly tried to keep up the interest levels of the students we were working with. This was generally done through interface changes. We made sure that we modify any interface element that causes a loss of perception, or is frustrating to the participants. This required iterative design and rapid prototyping.

### B. Technical challenges with Speech

Use of speech had a lot of attached technical challenges to it. As discussed earlier, male and female voices were hard to adapt to, but it was accomplished by using MLLR transforms for speaker adaptation. Speech recognition systems are generally very sensitive to background noise and environment changes; therefore we had to be careful about keeping the environment constant and stable across various sessions. We also used a high quality noise reduction microphone to capture audio during game play and during tests, to minimize effects of background noise.

### C. Administrative challenges

We also faced some administrative challenges, which we would wish to share with researchers who are working with similar communities in the developed world. Before starting the research, the IRB asked us for an approval for a study from a school. However, when we approached some potential schools, they asked us for an IRB approval. None of the agencies was at fault in this case, but it created a deadlock for us. However, having or developing contacts in a school administration generally helps in such cases. It also helps to develop good relations with local stakeholders like teachers. Our study was only possible because one of the ELL/ESL (English as a Second Language) teachers was excited to see our applications and tried hard to fit the experiment schedule into the class schedule, in a non-disruptive fashion. We also reached a consensus on the syllabus, with the teacher. This ensured that there were no confounding variables influencing the learning gains of the participants.

## XII. CONCLUSION AND FUTURE WORK

We hope that the positive results shown in this piece of research would inspire other such efforts in the field and underprivileged communities in the developed world would also start benefitting from the field of ICTD.

As stated earlier, the community we worked with did not necessarily have an intrinsic motivation to learn English. Therefore similar future ventures should involve efforts towards motivational games. Emotional analysis of speech can be used to detect the motivation levels or emotional states of the children. This information can then be used to start emotional conversations with the students. Games with conversational agents seem to be a good fit in such cases.

Our qualitative results point to some common player profiles that we observed during the game sessions. We would want to cater to all the profiles through our future games. There is a possibility of developing adaptive games, which try to gauge the profile of the player based on her interactions with the game and match that accordingly.

Moreover, we found that a "one size fits all" approach doesn't work in term of gender. Therefore, there is a need for games that have multiple story lines, characters, goals, reward structures and endings. In such cases interactive fiction seems like a good fit, where story, characters and plots could change based on the personality of the player. The kind of decision he/she takes in a game session would then determine the overall direction of the game. This would result in games that are still "one size fits all", but are considerate of gender, context and culture.

Our qualitative findings suggest that there was demand for multiplayer or collaborative games. Future research should try to explore implications of speech and games in the domain of shared learning. In such games, the players can collaborate and help each other with pronunciations.

Moreover, we would also want to work with younger children in the future. Human pronunciations are more susceptible to change for younger children [21]. We have already shown statistically significant gains for high school students, and we would want to test out these concepts with pre-school to middle school students too.

We would also want to conduct long-term experiments, with more hours of English instructions. We would also want to use sound pedagogical concepts to ensure retention. We would want to explore methodologies or practices that are specific to speech and pronunciation training.

The most important future work would be to look into the domain of mobile devices. With the increase in the processing power of the phones, it is possible to run speech recognizers on cellphones. We have already ported the CMU Sphinx-III speech recognition engine to mobile devices (Nokia N810), and the performance is comparable to the computer version (average time taken in decoding one word on Sphinx III is 0.92 seconds, and average time taken in decoding a word on the ported mobile version is 2.2 seconds).

We believe that with some or all of these changes incorporated into our next phase of research, we will be able to cause a greater change.

XIII. ACKNOWLEDGMENT

We would like to thank Mrs. Gilbert, Mrs. Sue Kayton and Mr. Steve Lippi for being our contacts at the test bed, and making we didn't face administrative issues. We would also want to thank Timothy Price, for his guidance with the linguistic aspects of the project. We are also immensely thankful to the students who were enrolled in our program. We also want to thank the reviewers for their valuable feedback, and Anuj Kumar for help with proofreading.